# Dimensionality, secondary flows and helicity in low-Rm MHD vortices

## Nathaniel T. Baker[1,2], Alban Pothérat[1] and Laurent Davoust[2]

[1]Coventry University, Applied Mathematics Research Center, Priory Street, Coventry CV15FB, UK

[2]Grenoble-INP / CNRS / Univ. Grenoble-Alpes, SIMaP, Electromagnetic Processing of Materials (EPM) Laboratory, F-38000 Grenoble, France

In this paper, we examine the dimensionality of a single electrically driven vortex bounded by two no-slip and perfectly insulating horizontal walls distant by $h$. The study was performed in the weakly inertial limit by means of an asymptotic expansion, which is valid for any Hartmann number. We show that the dimensionality of the leading order can be fully described using the single parameter $l_z^\nu/h$, where $l_z^\nu$ represents the distance over which the Lorentz force is able to act before being balanced by viscous dissipation. The base flow happens to introduce inertial recirculations in the meridional plane at the first order, which are shown to follow two radically different mechanisms: inverse Ekman pumping driven by a vertical pressure gradient along the axis of the vortex, or direct Ekman pumping driven by a radial pressure gradient in the Hartman boundary layers. We demonstrate that when the base flow is quasi-2D, the relative importance of direct and inverse pumping is solely determined by the aspect ratio $\eta/h$, where $\eta$ refers to the width of the vortex. Of both mechanisms, only inverse pumping appears to act as a significant source of helicity.

**Key words:**

---

## 1. Introduction

The question of the dimensionality of plane fluid layers is crucial to understand the mechanisms of energy dissipation in turbulence, which occurs in a number of natural and industrial problems. Indeed, whether three-dimensionality is present or not decides whether turbulence transfers energy to large, weakly dissipative structures (two-dimensional turbulence) or efficiently dissipates energy at small scales in the bulk of the flow (Tabeling (2002), Clercx & Van Heijst (2009)). These antagonistic behaviors may even co-exist in a number of real world situations. For instance, planetary atmospheres feature quasi two-dimensional (2D) large scales transporting small-scale three-dimensional (3D) turbulence (Lindborg 1999). Magnetohydrodynamic (MHD) turbulence in external magnetic fields (found in planetary cores or heat extracting devices) also share this feature (Klein & Pothérat 2010). Two-dimensionality can be lost when either a third velocity component or velocity gradients appear across the fluid layer. Shats *et al.* (2010) showed that the former mechanism accompanied the disruption of the inverse cascade in two-dimensional turbulence. Transversal transport across the layer is all the more important as it is often associated to strong vorticity and therefore contributes to generate kinematic helicity (Deusebio & Lindborg 2014). This quantity not only alters the properties of turbulence but also greatly helps to sustain dynamos (Gilbert *et al.* 1988).

Thanks to the diverse ways that exist to two-dimensionalize a flow (e.g. shallow confinement, background magnetic field or rotation), a broad range of studies focusing on elementary structures have flourished only to come up with the same conclusion: three dimensionality occurs as a result of gradients across the fluid layer. From a practical point of view, such gradients are usually introduced by no-slip boundaries, which are an intrinsic feature of any real system. A first attempt to quantify the quasi-2D / 3D structure of a single vortex confined in a shallow container was conducted by Satijn *et al.* (2001). They numerically studied the decay of a vortex, vertically bounded by a no-slip wall at the bottom and a free surface at the top, and showed that the relationship between shallow confinement and quasi-2D behavior was not straightforward. Furthermore they characterized the dimensionality of the vortex as a function of two non dimensional parameters characterizing the diffusion of momentum in the vertical and horizontal directions respectively. Later on, Akkermans *et al.* (2008) were able to visualize the recirculations that take place in front of a dipolar vortex traveling in a similar configuration as above. Their main conclusion was that three dimensionality resulted from vertical gradients of horizontal quantities (either velocity or forcing). It is only recently that the topological dimensionality of low-Rm MHD turbulence was elucidated by Pothérat & Klein (2014) in the light of diffusion lengths associated to the rotational part of a no-Lorentz force (first introduced by Sommeria & Moreau (1982)). A recent study from Pothérat *et al.* (2013), based on numerical simulations and experimental observations showed that meridional recirculations could either follow direct or inverse pumping.

There are however a number of limitations inherent to numerical and experimental methods that left the question in suspense since they could not investigate the finer properties of the flow. On the one hand, the main shortcoming of any numerical study comes from accessible computational power. Even though there was a recent breakthrough in solving low-Rm MHD turbulent flows in wall bounded domains (Kornet & Pothérat 2015), the regimes reachable by DNS are, to date, still far from those encountered experimentally. On the other hand, experiments are limited by the resolution of the measuring devices in use. These issues prevent a thorough investigation of the boundary layers, which happen to be one of the most crucial sources of three-dimensionality.

The work that we expose here takes place within the low-Rm MHD framework, in the continuity of the studies described above. The goal of this paper is to characterize the relationship between the topological dimensionality of a wall-bounded electrically driven vortex, and the occurring secondary flows. In order to do so, we focus on a single axisymmetric vortex confined between two horizontal planes. An analytical solution to the problem is derived by means of an asymptotic expansion in the weakly inertial limit, which is carried out up to the first order. The investigation performed here is valid for any Hartmann number, which makes it possible to investigate the finer properties of the boundary layers without concessions.

The model's geometry and governing equations are presented in section 2, and the exact solutions at leading and first order are given in sections 3 and 4 respectively. Section 5 is dedicated to the validation of the method, while section 6 is dedicated to presenting and discussing our results.

## 2. Geometry and governing equations

Let us consider an axisymmetric flow taking place in a cylindrical cavity of radius $R$. As such, we focus exclusively on solutions which are invariant to rotations about the axis of the channel i.e. $\partial_\theta = 0$. The domain is bounded by two no-slip horizontal walls located at $z = 0$ and $z = h$, and is filled with an electrically conducting fluid (typically a liquid metal such as Galinstan, of electrical conductivity $\sigma = 3.4 \times 10^6$ S/m, density $\rho = 6400$ kg/m³, and viscosity $\nu = 4 \times 10^{-7}$ m²/s). A static and uniform magnetic field $B_0\,\mathbf{e}_z$ is applied vertically. The low-Rm approximation is assumed to hold, meaning that the magnetic field induced by the flow is negligible compared to the imposed magnetic field (Roberts 1967). Consequently, the total magnetic field $\mathbf{B}$ is uniform across the domain and follows $\mathbf{B} = B_0\,\mathbf{e}_z$. In addition, the electric field $\mathbf{E}$ derives from the electric potential $\phi$ according to $\mathbf{E} = -\boldsymbol{\nabla}\phi$. A flow is driven by injecting electric current through an electrode of radius $\eta$ located on the bottom plate. The top and bottom plates are perfectly electrically insulating otherwise, which forces the current to exit the channel through the sides. In anticipation of the upcoming calculations, the profile of injected current is assumed to be a smooth function, such as the Gaussian distribution:

$$j_z^w(r) = \frac{I_0}{\pi\eta^2}\exp\left[-(r/\eta)^2\right],\tag{2.1}$$

where $I_0$ is the total current injected inside the domain up to the correction factor $\left(1 - \exp\left[-(R/\eta)^2\right]\right)$, which is almost identical to 1 for $R/\eta > 10$. Given this configuration, the electric current is known to flow radially, interacting with the vertical magnetic field to induce a patch of vertical vorticity right above the bottom Hartmann layer.

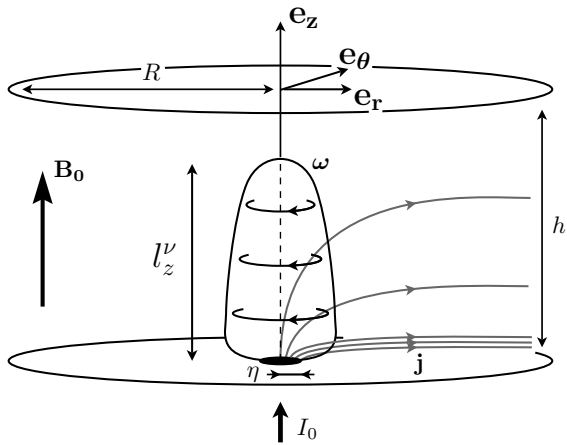

Figure 1: sketch of the problem. An isolated vortex of height $l_z^\nu < h$ confined between two horizontal no-slip and electrically insulating walls distant by $h$.

In the inertialess limit (Kalis & Kolesnikov 1980), the development of this patch of vorticity relies on the competition between two effects. On the one hand, the rotational part of the Lorentz force diffuses momentum along the magnetic field (Sommeria & Moreau 1982), hence leading to a vortex extending in the $z$ direction. On the other hand, viscous friction diffuses momentum isotropically, therefore opposing the growth of the vortex along $z$. Calling $l_z^\nu$ the range of action of the Lorentz force, its diffusive effect takes place over the characteristic time $\tau_{2D} = (\rho/\sigma B_0^2)\,(l_z^\nu/\eta)^2$. Conversely, viscous dissipation takes place over the time $\tau_\nu = \eta^2/\nu$. Assuming a steady flow, the distance $l_z^\nu$ over which the Lorentz force is able to act before being balanced by viscous dissipation is derived by equating both effects, yielding:

$$\frac{l_z^\nu}{h} = \frac{\eta^2}{h^2}\,Ha,\tag{2.2}$$

where $Ha = B_0 h \sqrt{\sigma/\rho\nu}$ is the Hartmann number based on the height of the channel. Asymptotically speaking, $l_z^\nu/h \ll 1$ means that the diffusive effect of the Lorentz force is balanced by viscous dissipation long before momentum can reach the top wall. In this case, the distance $l_z^\nu$ may be physically interpreted as the height of the vortex. On the contrary, $l_z^\nu/h \gg 1$ means that momentum can be diffused far beyond the top wall. This process is however blocked by the presence of the no-slip top wall, which prevents the vortex from extending past it. The ratio $l_z^\nu/h$ has been identified by Pothérat & Klein (2014) as the non dimensional parameter defining whether the structure is able to feel the presence of the top wall, hence controlling its dimensionality: 3D when $l_z^\nu/h \ll 1$ and quasi-2D when $l_z^\nu/h \gg 1$.

From now on, let us use the dimensionless coordinates $\tilde{r} = r/\eta$ and $\tilde{z} = z/h$, as well as the non dimensional variables $\tilde{u} = u/U$, $\tilde{\omega} = \omega\,\eta/U$, $\tilde{j} = j/\sigma U B_0$ and $\tilde{\phi} = \phi/U B_0 \eta$. We also introduce the non dimensional operator $\tilde{\boldsymbol{\nabla}}$ defined as

$$\tilde{\boldsymbol{\nabla}} = \left( \frac{\partial}{\partial \tilde{r}}, \, \frac{1}{\tilde{r}}\frac{\partial}{\partial \theta}, \, \frac{\eta}{h}\frac{\partial}{\partial \tilde{z}} \right).$$

The scaling for the velocity $U$ is derived from the linear theory of quasi-2D electrically driven vortices put forward by Sommeria (1988). It is estimated from $U = (\Gamma/\eta)\sqrt{l_z^\nu/h}$, where $\Gamma = I_0/2\pi\sqrt{\sigma\rho\nu}$ is the circulation induced right above a point-like electrode through which flows the current $I_0$, when viscous friction in the horizontal plane is neglected. This scaling for $U$ is representative of the velocity at the edge of the vortex core, whose radius results from the competition between the Lorentz force and viscous dissipation. Hence the explicit dependence of $U$ on the ratio $l_z^\nu/h$. The governing equations consist of the steady state vorticity equation for $\tilde{\omega} = \tilde{\boldsymbol{\nabla}} \times \tilde{\mathbf{u}}$

$$\frac{1}{N}\left( \tilde{\mathbf{u}}\cdot\tilde{\boldsymbol{\nabla}}\tilde{\boldsymbol{\omega}} - \tilde{\boldsymbol{\omega}}\cdot\tilde{\boldsymbol{\nabla}}\tilde{\mathbf{u}} \right) = \frac{1}{Ha}\left(\frac{l_z^\nu}{h}\right)^{-1}\tilde{\Delta}\tilde{\boldsymbol{\omega}} + \frac{1}{\sqrt{Ha}}\left(\frac{l_z^\nu}{h}\right)^{1/2}\frac{\partial\tilde{\mathbf{j}}}{\partial\tilde{z}}, \qquad (2.3)$$

Ohm's law

$$\tilde{\mathbf{j}} = -\tilde{\boldsymbol{\nabla}}\tilde{\phi} + \tilde{\mathbf{u}}\times\mathbf{e}_z, \qquad (2.4)$$

the conservation of mass

$$\tilde{\boldsymbol{\nabla}}\cdot\tilde{\mathbf{u}} = 0, \qquad (2.5)$$

and charge

$$\tilde{\boldsymbol{\nabla}}\cdot\tilde{\mathbf{j}} = 0. \qquad (2.6)$$

The problem at hand is governed by three non dimensional parameters, namely the interaction parameter $N$ based on the width of the injection electrode $\eta$, the Hartmann number $Ha$ based on the height of the channel $h$ and the ratio $l_z^\nu/h$:

$$N = \frac{\sigma B_0^2 \eta}{\rho U}, \quad Ha = B_0 h\sqrt{\frac{\sigma}{\rho\nu}}, \quad \frac{l_z^\nu}{h} = \frac{\eta^2}{h^2}\,Ha. \qquad (2.7)$$

The boundary conditions on the horizontal walls consist of no-slip boundaries

$$\tilde{\mathbf{u}}(\tilde{r},0) = \tilde{\mathbf{u}}(\tilde{r},1) = 0, \qquad (i)$$

an imposed vertical current at the bottom wall

$$\tilde{j}_{\tilde{z}}(\tilde{r},0) = \tilde{j}_{\tilde{z}}^w(\tilde{r}), \qquad (ii)$$

and a perfectly electrically insulating top wall

$$\tilde{j}_{\tilde{z}}(\tilde{r}, 1) = 0. \qquad (iii)$$

In addition, we impose a perfectly conducting and free slip radial boundary

$$\tilde{j}_{\tilde{z}}(\tilde{R}, \tilde{z}) = 0, \qquad (iv)$$

and

$$\tilde{\boldsymbol{\tau}}_{\tilde{r}}(\tilde{R}, \tilde{z}) = 0, \qquad (v)$$

where $\tilde{\boldsymbol{\tau}}_{\tilde{r}}$ represents the shear stress exerting on the wall whose normal vector is $\mathbf{e}_r$. These boundary conditions were chosen to match as closely as possible the experimental setups of (Sommeria 1988) and (Pothérat & Klein 2014), where a flow is driven by injecting a known amount of electric current $I_0$. The free-slip and perfectly conducting radial boundary can be physically interpreted as a pseudo-wall made of liquid metal, and was preferred over a no-slip boundary condition as it does not introduce parallel layers along the radial boundary. Keeping the experimental analogy in mind, the model we propose here focuses on an elementary structure, which has been extracted from an array of vortices.

Considering the scaling that was chosen for $U$, the normalized bottom boundary condition on the current $\tilde{j}_z^w$ is expressed as:

$$\tilde{j}_{\tilde{z}}^w(\tilde{r}) = \frac{2}{\sqrt{Ha}} \left(\frac{l_z^\nu}{h}\right)^{-1} \exp\left[-\dot{r}^2\right].$$

In other words, for a given value of $Ha$ and $l_z^\nu/h$, the intensity of the total injected current is adjusted so that the intensity of the resulting flow remains comparable throughout all cases investigated.

We shall now consider a weakly inertial flow in the limit $N \gg 1$, and expand equations (2.3) through (2.6) using the regular perturbation series:

$$\tilde{\mathbf{j}} = \tilde{\mathbf{j}}^0 + N^{-1}\tilde{\mathbf{j}}^1 + N^{-2}\tilde{\mathbf{j}}^2 + O(N^{-3}),$$
$$\tilde{\mathbf{u}} = \tilde{\mathbf{u}}^0 + N^{-1}\tilde{\mathbf{u}}^1 + N^{-2}\tilde{\mathbf{u}}^2 + O(N^{-3}),$$
$$\tilde{\boldsymbol{\omega}} = \tilde{\boldsymbol{\omega}}^0 + N^{-1}\tilde{\boldsymbol{\omega}}^1 + N^{-2}\tilde{\boldsymbol{\omega}}^2 + O(N^{-3}).$$

## 3. Inertialess base flow

The equations governing the problem at leading order are:

$$\tilde{\Delta}\tilde{\boldsymbol{\omega}}^0 = -\sqrt{Ha}\left(\frac{l_z^\nu}{h}\right)^{3/2} \frac{\partial \tilde{\mathbf{j}}^0}{\partial \tilde{z}}, \qquad (3.1)$$

$$\tilde{\mathbf{j}}^0 = -\tilde{\boldsymbol{\nabla}}\tilde{\phi}^0 + \tilde{\mathbf{u}}^0 \times \mathbf{e}_z, \qquad (3.2)$$

$$\tilde{\boldsymbol{\nabla}} \cdot \tilde{\mathbf{u}}^0 = 0, \qquad (3.3)$$

$$\tilde{\boldsymbol{\nabla}} \cdot \tilde{\mathbf{j}}^0 = 0. \qquad (3.4)$$

Solving $\tilde{\boldsymbol{\omega}}^0$ and $\tilde{\mathbf{j}}^0$ can be done separately by taking the Laplacian of the vorticity equation (2.3) on the one hand, and taking the Laplacian of twice the curl of Ohm's law (2.4) on the other hand. Combining both of them yields the following set of equations:

$$\tilde{\Delta}^2\tilde{\boldsymbol{\omega}}^0 = \left(\frac{l_z^\nu}{h}\right)^2 \frac{\partial^2 \tilde{\boldsymbol{\omega}}^0}{\partial \tilde{z}^2}, \qquad (3.5)$$

and

$$\tilde{\Delta}^2 \tilde{\mathbf{j}}^0 = \left(\frac{l_z^\nu}{h}\right)^2 \frac{\partial^2 \tilde{\mathbf{j}}^0}{\partial \tilde{z}^2} \,. \tag{3.6}$$

It is quite remarkable that equations (3.5) and (3.6) can be solved independently, and depend on the same and unique parameter $l_z^\nu/h$. They remain nonetheless coupled via twice the curl of Ohm's law:

$$\tilde{\Delta} \tilde{\mathbf{j}}^0 = -\frac{1}{\sqrt{Ha}} \left(\frac{l_z^\nu}{h}\right)^{1/2} \frac{\partial \tilde{\boldsymbol{\omega}}^0}{\partial \tilde{z}} \,. \tag{3.7}$$

Equations (3.5) and (3.6) admit a purely azimuthal solution for $\tilde{\mathbf{u}}^0$ and a purely meridional solution for $\mathbf{j}^0$. Consequently, knowing either component $\tilde{\omega}_{\tilde{z}}^0$ or $\tilde{\omega}_{\tilde{r}}^0$, and $\tilde{j}_{\tilde{r}}^0$ or $\tilde{j}_{\tilde{z}}^0$ is enough to completely derive the solution at leading order. The boundary conditions associated to the leading order read:

$$\tilde{\omega}_{\tilde{z}}^0(\tilde{r}, 0) \;=\; \tilde{\omega}_{\tilde{z}}^0(\tilde{r}, 1) \;=\; 0, \tag{$i^0$}$$

$$\tilde{j}_{\tilde{z}}^0(\tilde{r}, 0) \;=\; \tilde{j}_{\tilde{z}}^w(\tilde{r}), \tag{$ii^0$}$$

$$\tilde{j}_{\tilde{z}}^0(\tilde{r}, 1) \;=\; 0, \tag{$iii^0$}$$

$$\tilde{j}_{\tilde{z}}^0(\tilde{R}, \tilde{z}) \;=\; 0. \tag{$iv^0$}$$

In addition, we shall approximate boundary condition $(v)$ by

$$\tilde{\omega}_{\tilde{z}}^0(\tilde{R}, \tilde{z}) \;=\; 0. \tag{$v^0$}$$

Boundary condition $(v^0)$ is not entirely equivalent to the free-slip boundary condition $(v)$, which can be re-written in terms of $\tilde{\omega}_{\tilde{z}}^0$ as:

$$\tilde{\omega}_{\tilde{z}}^0(\tilde{R}, \tilde{z}) = \frac{2}{\tilde{R}} \, \tilde{u}_\theta^0(\tilde{R}, \tilde{z}) \,. \tag{3.8}$$

It is only in the limit $\tilde{R} \gg 1$ (since $\tilde{u}_\theta^0$ is of order 1), that boundary condition $(v)$ may be approximated by $(v^0)$. However, $(v^0)$ offers a much simpler numerical implementation. Indeed, not only does it remove any remaining dependency on $\tilde{u}_\theta^0$, it also naturally introduces an orthogonal basis of functions on which the solution can be projected. From a practical point of view, we ensured that the edge of the channel was sufficiently far from the injection area in order to minimize the impact of this approximation on the flow (see section 5).

Solutions with separated variables of (3.5) and (3.6), which satisfy the coupling (3.7), as well as boundary conditions $(iv^0)$ and $(v^0)$ are of form:

$$\tilde{\omega}_{\tilde{z}}^0 = \sum_{n=1}^{\infty} J_0\left(\lambda_n \tilde{r}\right) \sum_{i=1}^{4} A_{ni} \exp\left(s_{ni} \tilde{z}\right), \tag{3.9}$$

and

$$\tilde{j}_{\tilde{z}}^0 = \sum_{n=1}^{\infty} J_0\left(\lambda_n \tilde{r}\right) \sum_{i=1}^{4} B_{ni} \exp\left(s_{ni} \tilde{z}\right), \tag{3.10}$$

where $J_0(\tilde{r})$ refers to the zeroth order Bessel function of first kind, and $\lambda_n$ represents its $n^{th}$ root normalized by $\tilde{R}$. Note that solutions with separated variables which satisfy the coupling (3.7) and boundary condition $(v^0)$ alone automatically satisfy boundary condition $(iv^0)$, making the latter redundant and therefore unnecessary to close the problem. Conversely, a different set of boundary conditions at $\tilde{r} = \tilde{R}$ may not admit a solution with separated variables. The arguments for the exponentials $s_{ni}$ may be expressed in terms of $Ha$ and $l_z^\nu/h$ exclusively. They take four different values $s_{ni} = \pm s_{n\pm}$, where $s_{n\pm}$ is defined by:

$$s_{n\pm} = \frac{Ha}{2} \left[ 1 \pm \sqrt{1 + \frac{4\,\lambda_n^2}{Ha} \left(\frac{l_z^\nu}{h}\right)^{-1}} \right]. \qquad (3.11)$$

Restricting ourselves to cases that are relevant to MHD (i.e. cases where $Ha$ is sufficiently large), the parameter $Ha^{-1} \left(l_z^\nu/h\right)^{-1}$ is expected to be much less than 1. Under this assumption, the roots $s_{n\pm}$ are expected to scale as

$$s_{n+} \ \sim \ \frac{h}{\delta} \qquad \text{and} \qquad s_{n-} \ \sim \ \lambda_n^2 \frac{h}{l_z^\nu}, \qquad (3.12)$$

where $\delta = h/Ha$ represents the thickness of the Hartmann boundary layer. That is to say, $s_{n+}$ scales as $1/\delta$, thus describes the boundary layers, while $s_{n-}$ scales as $\lambda_n^2/l_z^\nu$, which is the diffusion length associated to Bessel mode $n$. In this sense it represents the dimensionality of the bulk of the flow. From (3.7), coefficients $A_{ni}$ and $B_{ni}$ must satisfy:

$$B_{ni} = - A_{ni} \ \frac{s_{ni}}{s_{ni}^2 \kappa - \lambda_n^2/\kappa}, \qquad (3.13)$$

with $\kappa = Ha^{-1/2} \left(l_z^\nu/h\right)^{1/2}$. The coefficients $A_{ni}$ are determined by solving the linear system stemming from the boundary conditions:

$$\sum_{i=1}^{4} A_{ni} = 0,$$

$$\sum_{i=1}^{4} A_{ni} \, \exp\left(s_{ni}\right) = 0,$$

$$\sum_{i=1}^{4} A_{ni} \ \frac{s_{ni}}{s_{ni}^2 \kappa - \lambda_n^2/\kappa} = -\alpha_n, \qquad (S^0)$$

$$\sum_{i=1}^{4} A_{ni} \ \frac{s_{ni}}{s_{ni}^2 \kappa - \lambda_n^2/\kappa} \, \exp\left(s_{ni}\right) = 0,$$

where $\alpha_n$ results from the projection of $\tilde{j}_z^w(\tilde{r})$ on the basis of Bessel functions:

$$\alpha_n = \frac{2/\tilde{R}^2}{J_1^2(\lambda_n\,\tilde{R})} \int_0^{\tilde{R}} \xi \, \tilde{j}_z^w(\xi) \, J_0(\lambda_n\,\xi) \, d\xi. \qquad (3.14)$$

At this stage, the supplementary radial components $\tilde{\omega}_r^0$ and $\tilde{j}_r^0$, as well as the velocity field $\tilde{\mathbf{u}}^0 = \tilde{u}_\theta^0 \, \mathbf{e}_\theta$, can be readily determined by integrating $\tilde{\boldsymbol{\nabla}} \cdot \tilde{\boldsymbol{\omega}}^0 = 0$, $\tilde{\boldsymbol{\nabla}} \cdot \tilde{\mathbf{j}}^0 = 0$, and $\tilde{\boldsymbol{\omega}}^0 = \tilde{\boldsymbol{\nabla}} \times \tilde{\mathbf{u}}^0$ respectively.

## 4. Correction due to inertia

The equations governing the problem at first order are:

$$\tilde{\mathbf{u}}^0 \cdot \tilde{\boldsymbol{\nabla}} \tilde{\boldsymbol{\omega}}^0 - \tilde{\boldsymbol{\omega}}^0 \cdot \tilde{\boldsymbol{\nabla}} \tilde{\mathbf{u}}^0 = \frac{1}{Ha} \left(\frac{l_z^\nu}{h}\right)^{-1} \tilde{\Delta} \tilde{\boldsymbol{\omega}}^1 + \frac{1}{\sqrt{Ha}} \left(\frac{l_z^\nu}{h}\right)^{1/2} \frac{\partial \tilde{\mathbf{j}}^1}{\partial \tilde{z}}, \qquad (4.1)$$

$$\tilde{\mathbf{j}}^1 = -\tilde{\boldsymbol{\nabla}} \tilde{\phi}^1 + \tilde{\mathbf{u}}^1 \times \mathbf{e}_z, \qquad (4.2)$$

$$\tilde{\boldsymbol{\nabla}} \cdot \tilde{\mathbf{u}}^1 = 0 \,, \tag{4.3}$$

$$\tilde{\boldsymbol{\nabla}} \cdot \tilde{\mathbf{j}}^1 = 0 \,. \tag{4.4}$$

Unlike the leading order (which is forced electrically at the bottom wall), the first order is driven by an azimuthal inertial force stemming from the base flow. In other words, $\tilde{\omega}_{\tilde{z}}^1$ and $\tilde{j}_{\tilde{z}}^1$ must satisfy homogeneous boundary conditions all along the edges of the domain. As a consequence, $\tilde{\omega}_{\tilde{z}}^1$ is strictly null, and $\tilde{\phi}^1$ is uniform across the channel. In order to have a non-divergent solution on the axis of the channel, $\tilde{\omega}_{\tilde{r}}^1$ must also be null throughout the domain, meaning that the inertial correction to the base flow occurs in the meridional plane exclusively. In addition, the electric current becomes purely electromotive, since it is proportional to the velocity via Ohm's law. These arguments simplify the problem greatly by removing all couplings between mechanical and electrical quantities at first order. In the end, the equations reduce to:

$$\tilde{F}_\theta^0 \, \mathbf{e}_\theta = \frac{1}{Ha} \left(\frac{l_{\tilde{z}}^\nu}{h}\right)^{-1} \tilde{\Delta} \tilde{\boldsymbol{\omega}}^1 - \frac{1}{\sqrt{Ha}} \left(\frac{l_{\tilde{z}}^\nu}{h}\right)^{1/2} \frac{\partial \tilde{u}_{\tilde{r}}^1}{\partial \tilde{z}} \mathbf{e}_\theta, \tag{4.5}$$

where $\tilde{F}_\theta^0$ is the inertial forcing originating from the non linear terms of the base flow:

$$\tilde{F}_\theta^0 = 2 \, \frac{\tilde{u}_\theta^0 \, \tilde{\omega}_{\tilde{r}}^0}{\tilde{r}}. \tag{4.6}$$

Owing to the previous arguments, equation (4.5) is non trivial only in the $\mathbf{e}_\theta$ direction. It is solved by introducing the stream function $\tilde{\psi}^1 = \psi^1/U\eta$ such that $\tilde{\mathbf{u}}^1 = \tilde{\boldsymbol{\nabla}} \times (\tilde{\psi}^1 \, \mathbf{e}_\theta)$ and $\tilde{\boldsymbol{\omega}}^1 = \tilde{\boldsymbol{\nabla}} \times \tilde{\boldsymbol{\nabla}} \times (\tilde{\psi}^1 \, \mathbf{e}_\theta)$, yielding the following equation for $\tilde{\psi}^1$:

$$\frac{l_{\tilde{z}}^\nu}{h} Ha \, \tilde{F}_\theta^0 = - \left[ \tilde{\Delta} - \frac{1}{\tilde{r}^2} \right]^2 \tilde{\psi}^1 + \left(\frac{l_{\tilde{z}}^\nu}{h}\right)^2 \frac{\partial^2 \tilde{\psi}^1}{\partial \tilde{z}^2}, \tag{4.7}$$

where $\tilde{\Delta}$. represents the scalar Laplace operator in cylindrical coordinates. Again, the intensity of the flow depends on the interaction parameter $N$, while the topology of the first order recirculations depends on $Ha$ and the ratio $l_{\tilde{z}}^\nu/h$. Since mechanical and electrical quantities are decoupled at first order, the boundary conditions associated to this problem boil down to no-slip and non-penetrating boundaries at the top and bottom walls, and no shear stress at the radial boundary:

$$\left.\frac{\partial \tilde{\psi}^1}{\partial \tilde{z}}\right|_{\tilde{r},0} = \left.\frac{\partial (\tilde{r}\tilde{\psi}^1)}{\partial \tilde{r}}\right|_{\tilde{r},0} = 0, \tag{$i_0^1$}$$

$$\left.\frac{\partial \tilde{\psi}^1}{\partial \tilde{z}}\right|_{\tilde{r},1} = \left.\frac{\partial (\tilde{r}\tilde{\psi}^1)}{\partial \tilde{r}}\right|_{\tilde{r},1} = 0, \tag{$i_1^1$}$$

$$\left.\frac{\partial}{\partial \tilde{r}} \left( \frac{1}{\tilde{r}} \frac{\partial (\tilde{r}\tilde{\psi}^1)}{\partial \tilde{r}} \right)\right|_{\tilde{R},\tilde{z}} - \left.\frac{\partial^2 \tilde{\psi}^1}{\partial \tilde{z}^2}\right|_{\tilde{R},\tilde{z}} = 0. \tag{$v^1$}$$

Solutions of equation (4.7) are sought for as the sum of a homogeneous solution $\tilde{\psi}_h^1(\tilde{r}, \tilde{z})$, and a particular solution of the problem with inertial forcing $\tilde{\psi}_f^1(\tilde{r}, \tilde{z})$. The homogeneous problem is very similar to the one solved earlier and reads:

$$\tilde{\psi}_h^1(\tilde{r}, \tilde{z}) = \sum_{n=1}^{\infty} J_1(\mu_n \, \tilde{r}) \sum_{i=1}^{4} C_{ni} \, \exp\left(p_{ni} \, \tilde{z}\right), \tag{4.8}$$

where $J_1(\tilde{r})$ is the first order Bessel function of first kind, and $\mu_n$ represents its $n^{th}$ root normalized by $\tilde{R}$. As for the leading order, the roots $p_{ni}$ are defined as

$$p_{n\pm} = \frac{Ha}{2} \left[ 1 \pm \sqrt{1 + \frac{4\,\mu_n^2}{Ha} \left(\frac{l_z^\nu}{h}\right)^{-1}} \right]. \tag{4.9}$$

The particular solution $\psi_f^1$ is found by expanding $\tilde{F}_\theta^0$ as a Fourier-Bessel series of $J_1(\mu_n\,\tilde{r})$:

$$\tilde{F}_\theta^0(\tilde{r}, \tilde{z}) = \sum_{n=1}^\infty J_1(\mu_n\,\tilde{r})\,\varphi_n(\tilde{z}), \tag{4.10}$$

with

$$\varphi_n(\tilde{z}) = -2\,\kappa \sum_{i,j=1}^\infty \sum_{k,l=1}^4 \beta_{nij}\,\frac{A_{ik}\,A_{jl}\,s_{jl}}{\lambda_i\,\lambda_j}\,\exp\left[(s_{ik} + s_{jl})\,\tilde{z}\right], \tag{4.11}$$

where

$$\beta_{nij} = \frac{2/\tilde{R}^2}{J_2^2(\mu_n\,\tilde{R})} \int_0^{\tilde{R}} \xi\,J_1(\lambda_i\,\xi)\,J_1(\lambda_j\,\xi)\,J_1(\mu_n\,\xi)\,d\xi. \tag{4.12}$$

The response of the flow to the forcing is therefore

$$\tilde{\psi}_f^1(\tilde{r}, \tilde{z}) = \sum_{n=1}^\infty J_1(\mu_n\,\tilde{r}) \sum_{i,j,k,l} K_{nijkl} \exp\left[(s_{ik} + s_{jl})\,\tilde{z}\right], \tag{4.13}$$

where

$$K_{nijkl} = \frac{\beta_{nij}\,\dfrac{2}{\kappa}\,\left(\dfrac{l_z^\nu}{h}\right)^2\,\dfrac{A_{ik}\,A_{jl}\,s_{jl}}{\lambda_i\,\lambda_j}}{\mu_n^4 - \left[\left(\dfrac{l_z^\nu}{h}\right)^2 + 2\,(\mu_n\,\kappa)^2\right](s_{ik} + s_{jl})^2 + \left[\kappa\,(s_{ik} + s_{jl})\right]^4}. \tag{4.14}$$

Note that at this stage the value of $K_{nijkl}$ is fully determined, since it only depends on quantities resulting from the base flow. Finally,

$$\tilde{\psi}^1 = \sum_{n=1}^\infty J_1(\mu_n\,\tilde{r}) \left\{ \sum_{i=1}^4 C_{ni} \exp(p_{ni}\,\tilde{z}) + \sum_{i,j,k,l} K_{nijkl} \exp\left[(s_{ik} + s_{jl})\,\tilde{z}\right] \right\}, \tag{4.15}$$

where the integration constants $C_{ni}$ are determined from the boundary conditions:

$$\begin{aligned} \sum_{i=1}^4 C_{ni} &= -\sum_{i,j,k,l} K_{nijkl}, \\ \sum_{i=1}^4 C_{ni}\,p_{ni} &= -\sum_{i,j,k,l} K_{nijkl}\,(s_{ik} + s_{jl}), \\ \sum_{i=1}^4 C_{ni}\exp(p_{ni}) &= -\sum_{i,j,k,l} K_{nijkl}\exp(s_{ik} + s_{jl}), \\ \sum_{i=1}^4 C_{ni}\,p_{ni}\exp(p_{ni}) &= -\sum_{i,j,k,l} K_{nijkl}\,(s_{ik} + s_{jl})\exp(s_{ik} + s_{jl}). \end{aligned} \tag{$S^1$}$$

To summarize, equations (3.9), (3.10) and (4.15) provide a complete solution for the flow at order $O(N^{-1})$ in the limit $N \gg 1$, and for any arbitrary value of $Ha$ or $l_z^\nu/h$. This solution is shown to be exclusively governed by three non dimensional parameters: $N$, which determines the intensity of the flow (as in the theory of Pothérat *et al.* (2000)); $Ha$, which controls the thickness of the boundary layers (as in the classical theory); and $l_z^\nu/h$, which defines the dimensionality of the flow. With such a formulation of the problem, one can clearly see that the geometrical aspect ratio $\eta/h$ is not the most adequate parameter to precisely describe the dimensionality of the base flow. However $l_z^\nu/h$ is. This may explain why the height-to-width aspect ratio of the vortex can be seen as a "confusing parameter" (Satijn *et al.* 2001). It is also worth noting that the even orders correspond to the azimuthal component of the flow, while the odd orders give corrections in the meridional plane. This behavior was also found in the analogous configuration described by Davoust *et al.* (2015), which consists of an annular channel with a rotating bottom. We shall now numerically evaluate this solution to find how the topological dimensionality of the base flow impacts the secondary recirculations.

## 5. Numerical methods

### 5.1. *Algorithm description*

An in-house FORTRAN95 code was developed to evaluate numerically $\tilde{\omega}_{\tilde{z}}^0(\tilde{r}, \tilde{z})$, $\tilde{j}_{\tilde{z}}^0(\tilde{r}, \tilde{z})$ and $\tilde{\psi}^1(\tilde{r}, \tilde{z})$. The goal of the solver was to compute the Fourier-Bessel coefficients $A_{ni}$ and $C_{ni}$. That is to say, solve systems $(S^0)$ and $(S^1)$ in order to re-construct the solution via equations (3.9), (3.10) and (4.15) respectively. The FMLIB 1.3 multi-precision package (Smith 1991) was used to ensure sufficient accuracy of the solution for any value of $Ha$. The input parameters for our code were $Ha$, $l_z^\nu/h$, and the number of modes $N_{\text{mode}}$. The structure of the algorithm is as follows.

(*a*)  Set the working precision based on $Ha$

(*b*)  Generate the zeros of Bessel functions $J_0$ and $J_1$.

(*c*)  Compute $\alpha_n$ and $\beta_{nij}$ by evaluating the integrals using a Gauss-Legendre quadrature rule of order 100.

(*d*)  Compute $s_{n\pm}$ and $p_{n\pm}$ according to (3.11) and (4.9) respectively.

(*e*)  Compute $K_{nijkl}$ according to (4.14).

(*f*)  Find $A_{ni}$ and $C_{ni}$ by solving $(S^0)$ and $(S^1)$ with a Gauss-Jordan elimination method

(*g*)  Discretize domain, build the solution according to (3.9), (3.10) and (4.15), and convert output to double precision.

### 5.2. *Convergence test*

In this section, we evaluate the number of terms necessary to accurately represent infinite sums. To this end, let us introduce $\epsilon^n$, the relative error at order $n$ respectively defined by

$$\epsilon^0(N_{mode}) = \frac{\left\| \tilde{u}_\theta^0(N_{mode}) - \tilde{u}_\theta^0(N_{max}) \right\|_2}{\left\| \tilde{u}_\theta^0(N_{max}) \right\|_2}, \tag{5.1}$$

and

$$\epsilon^1(N_{mode}) = \frac{\left\| \tilde{\psi}^1(N_{mode}) - \tilde{\psi}^1(N_{max}) \right\|_2}{\left\| \tilde{\psi}^1(N_{max}) \right\|_2}, \tag{5.2}$$

where $\| \cdot \|_2$ represents the classically defined $\mathcal{L}^2$-norm. $\epsilon^n$ compares the difference between a run computed with the number of modes $N_{\text{mode}}$, with respect to the reference case computed with the highest number of modes $N_{\text{max}}$. For all cases, $N_{\text{max}}$ was set to 80. The convergence tests were conducted for three different channel radii $\tilde{R} = 5$, 10 and 20, since this parameter was expected to impact the accuracy of the solution. According to figure 2, the number of modes required to achieve a given relative error unsurprisingly increases with $\tilde{R}$. Indeed, the vortex becomes thinner with respect to the total width of the channel, meaning that modes of smaller wavelengths are required to capture it precisely.

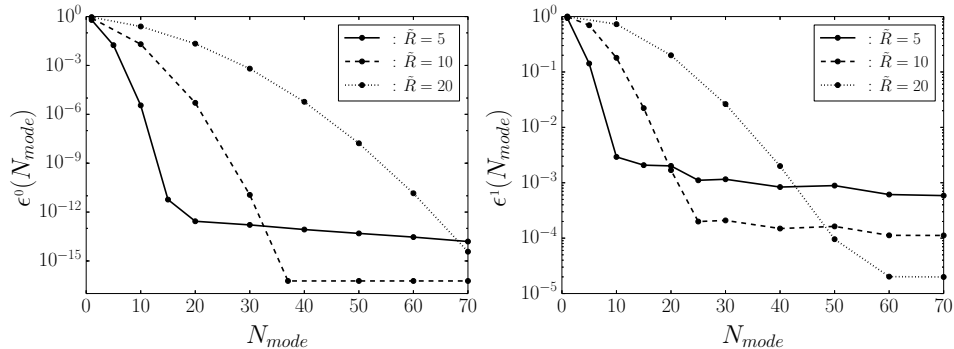

Figure 2: convergence test. Left: leading order; Right: first order.

At leading order, increasing the number of modes with $\tilde{R} = 10$ and $\tilde{R} = 20$ steadily improves the accuracy of the solution until $\epsilon^0$ eventually reaches double precision. The behavior of $\epsilon^0$ for $\tilde{R} = 5$ is completely different: fast convergence is observed at first, followed by a region where accuracy hardly improves with $N_{\mathrm{mode}}$. This effect is first evidence that the radial wall is too intrusive for $\tilde{R} = 5$.

At first order, $\epsilon^1$ follows a similar behavior regardless of the position of the radial wall: increasing the number of modes improves the residual error before it levels off. This behavior comes from the discrepancy that exists between the *real* inertial forcing $F_\theta^o$ (which is only approximately null at the edge due to the simplified boundary condition) and its Fourier-Bessel expansion, which is strictly null by definition of $J_1(\mu_n \tilde{R})$. This discrepancy introduces Gibbs phenomena close to the edge of the channel. It is however important to note that the oscillations are confined to a region close to the edge. Additionally, they become less and less an issue as $R$ is increased, since $F_\theta^o$ naturally vanishes away from the core of the vortex.

The conclusion of this convergence analysis is that $\tilde{R}$ must be as large as possible to prevent numerical artifacts. The operating point chosen was $\tilde{R} = 20$ and $N_{\mathrm{mode}} = 50$, which gave us a good compromise between accuracy and computational time (proportional to $N_{\mathrm{mode}}^3$). With these settings, the solution at leading order is reliable up to eight significant digits, and the relative accuracy of the first order is better than 0.01%

### 5.3. Validity of the radial boundary condition

Figure 3 shows the radial profiles of azimuthal velocity evaluated at the middle of the channel for different radial wall distances. In the case at hand, $l_z^\nu/h = 1000$, meaning that the base flow is already quasi-2D. According to figure 3, the azimuthal velocity follows a $1/\tilde{r}$ decay law outside the core of the vortex. This behavior is in agreement with the quasi-2D theory developed by Sommeria (1988) for a vortex driven by injecting current through a pointlike electrode. This suggests that the radial distribution of injected current plays a minor role in determining the actual shape of the vortex, and that a gaussian distribution provides a very good representation of a thin current injection electrode (at least when the flow is quasi two-dimensional). This point will be further studied in the following section.

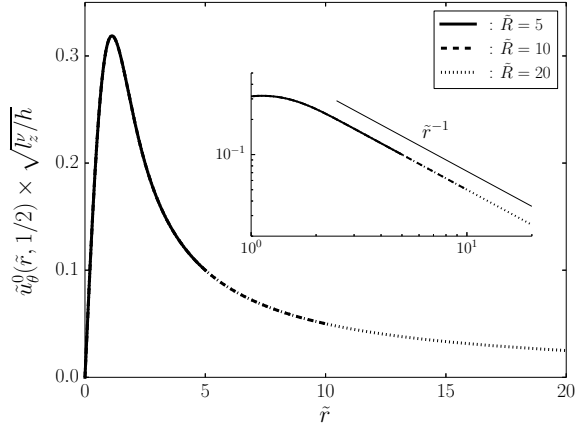

Figure 3: velocity profile at the middle of the channel, obtained for $Ha = 456$ and $l_z^\nu/h = 1000$. The insert highlights how the azimuthal velocity decays as $1/\tilde{r}$ away from the core of the vortex.

|  | $\tilde{R} = 5$ | $\tilde{R} = 10$ | $\tilde{R} = 20$ |
|---|---|---|---|
| $\tilde{\tau}_{r\theta}^0$ | $3.99 \times 10^{-2}$ | $1.00 \times 10^{-2}$ | $2.50 \times 10^{-3}$ |

Table 1: shear stress at the radial boundary

Table 1 gives an estimation of the leading order shear stress at the edge of the domain

$$\tilde{\tau}_{r\theta}^0 = \tilde{r} \, \frac{\partial}{\partial \tilde{r}} \left( \frac{\tilde{u}_\theta^0}{\tilde{r}} \right) \bigg|_{\tilde{R}, \tilde{z}}, \tag{5.3}$$

for different positions of the radial wall. It gives an a posteriori confirmation that the simplified radial boundary condition $(v^0)$ tends towards a free slip boundary condition when $\tilde{R}$ is increased. Furthermore, we find that $\tilde{\tau}_{r\theta}^0$ scales as $1/\tilde{R}^2$ to a very good precision. This brings supplementary evidence that the solution is reliable, since $\tilde{u}_\theta^0(\tilde{R})$ is expected to scale as $1/\tilde{R}$ for quasi-2D structures. In 3D flows, $\tilde{u}_\theta^0(\tilde{R})$ is expected to be lower, and so should be $\tilde{\tau}_{r\theta}^0$. For $\tilde{R} = 20$, the order of magnitude of the shear stress at the wall is of $10^{-3}$.

### 5.4. *Sensitivity to the injection profile and relevance to experiments*

Let us now investigate the sensitivity of the base flow to the bottom electric boundary condition. This question is all the more legitimate, as the very existence of the flow relies on the injection of electric current at the bottom. The spatial distribution of current density can thus be expected to shape the resulting flow. In order to quantify the relevance of our model to describe electrically driven vortices, we shall compare the flow induced by two different injection profiles:

$$j_z^w(r) = \frac{I_0}{\pi \eta^2} \exp \left[ -(r/\eta)^2 \right],$$

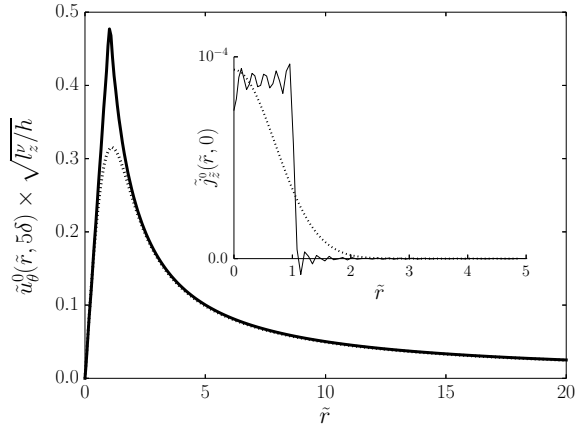

Figure 4: velocity profile right above the Hartmann layer induced by two different radial distributions of electric current. $\cdots$: flow resulting from a Gaussian distribution of current. ——: flow resulting from a step distribution of current. The insert shows the respective current profiles at the bottom wall $\tilde{j}_z^0(\tilde{r}, 0)$.

and

$$j_z^w(r) = \frac{I_0}{\pi \eta^2} \left[ H(r) - H(r - \eta) \right],$$

where $H(r)$ refers to the Heaviside step function. These two particular profiles were chosen so that the typical width of the electrode remained $\eta$, and that the total amount of electric current injected in the domain was $I_0$. For both cases $Ha$ and $l_z^\nu/h$ were set to $Ha = 456$ and $l_z^\nu/h = 1000$ respectively. Furthermore, the number of modes used to expand the Gaussian distribution was $N_{mode} = 50$ (in agreement with section 5.2), while $N_{mode} = 200$ was imposed to expand the step distribution. A much higher number of modes is obviously necessary for the latter profile since it is singular at $\tilde{r} = 1$.

Figure 4 shows the leading order azimuthal velocity along $\tilde{r}$, right above the bottom Hartmann layer ($\tilde{z} = 5/Ha$) for both current distributions. The associated current profiles are displayed in the insert. The first striking feature of figure 4 is that both velocity profiles follow the same asymptotic behavior whether close to the axis of the vortex or away from its core. This behavior comes from the fact that the lateral diffusion of momentum is driven by viscous dissipation, and is therefore independent of the injected electric current. As already discussed in section 5.3, both vortices follow a $1/\tilde{r}$ decay law away from their core, which is expected for quasi-2D structures. The velocity peak is found at $\tilde{r} = 1$ in both cases, which corresponds to the outer edge of the electrode. The main difference between both profiles however, is the value of the peak which is approximately twice as large for the step distribution. As a result, we can expect the Gaussian distribution to slightly underestimate the magnitude of the inertial terms. However, since the shape of the flow is identical in both cases, the mechanisms driving the first order recirculations will be unchanged (recall that the inertial forcing stems from mechanical quantities only).

As in experiments on electrically driven flows, the intensity of the vortex is controlled by the total imposed current through the electrode (Sommeria (1988), Messadek & Moreau (2002), Klein & Pothérat (2010), Pothérat & Klein (2014)). Nevertheless, it is also possible to impose a fixed voltage between electrodes, or between the injection electrode and the side wall. Kalis & Kolesnikov (1980) showed that imposing a uniform current density or uniform voltage at the electrode was essentially equivalent as far as the topology of the base flow was concerned. We can therefore assert that our model is a faithful representation of electrically driven vortices in experiments, even if a Gaussian distribution of current is imposed at the bottom.

It is also worth noting that although high frequency oscillations exist in the expansion of the step distribution (such oscillations are unavoidable regardless of the number of modes taken into account, as a result of its singular nature), they do not appear in the induced flow. This effect comes from the analytical approach that was used, and more specifically from the systematic use of dot products to build the solution. This brings supplementary evidence that our model is reliable and robust, since it is insensitive to numerical artifacts.

## 6. Results

Numerical experiments were conducted for four values of the Hartman number: $Ha = 456$, 911, 1822 and 3644. For all values of $Ha$, the ratio $l_z^\nu/h$ was set within the range $10^{-2}$ to $10^5$. From now on, the low-$Ha$ case refers to $Ha = 456$, while the high-$Ha$ case refers to $Ha = 3644$.

### 6.1. Inertialess base flow

Figure 5 depicts the base flow for $Ha = 3644$, and for three values of $l_z^\nu/h$: 0.01, 1 and 10000 respectively. The lower $Ha$ cases are not presented here, since they look almost identical. As a matter of fact, the only difference at leading order between them is the thickness of the boundary layers.

The dimensionality of the flow can be estimated by comparing the intensity of the velocity field along the top and bottom walls. For $l_z^\nu/h = 0.01$, the flow is mostly concentrated at the bottom of the channel (i.e. where the electric forcing takes place), while there is absolutely no flow at the top. The base flow is said to be weakly 3D (in the sense of Klein & Pothérat (2010)), meaning that although the topological patterns remain the same across the channel (that is to say, the vortex stays columnar), their intensity still depends on $\tilde{z}$. This behavior is a consequence of the Lorentz force not being strong enough to compete with viscous dissipation beyond $l_z^\nu$, which is in this case a hundred times smaller than $h$. In other words, weak three-dimensionality characterizes a flow where two adjacently stacked layers of horizontal velocity experience differential rotation, as a result of vertical gradients. As $l_z^\nu/h$ increases, the range of action of the Lorentz force becomes longer, and momentum is diffused further up the channel. For $l_z^\nu/h = 10000$, the flow is quasi-2D in the sense that all velocity gradients along $\tilde{z}$ have been smoothed out outside of the boundary layers (a $\tilde{z}$-dependence always exists in the top and bottom Hartmann layers due to the no-slip walls). The vortex spans across the channel and is therefore able to feel the effect of the top wall.

Figure 5 also displays current densities. As expected, they are highest where strong velocity gradients exist, i.e. in the boundary layers and in the core of the vortex. For all cases, we have verified that up to numerical precision, exactly half of the total current injected in the channel flows within the bottom Hartmann layer, while the other half flows vertically. This result confirms the heuristic prediction of Pothérat & Klein (2014). For low values of $l_z^\nu/h$ (0.01 and 1), the velocity gradient along $\tilde{z}$ introduced by three-dimensionality progressively extracts the vertical current into the bulk, channeling it towards the edge of the domain. For $l_z^\nu/h = 10000$ however, quasi two-dimensionality has smoothed out all velocity gradients along $\tilde{z}$ in the bulk: the vertical current flows exclusively within the core of the vortex and the top and bottom Hartmann layers.

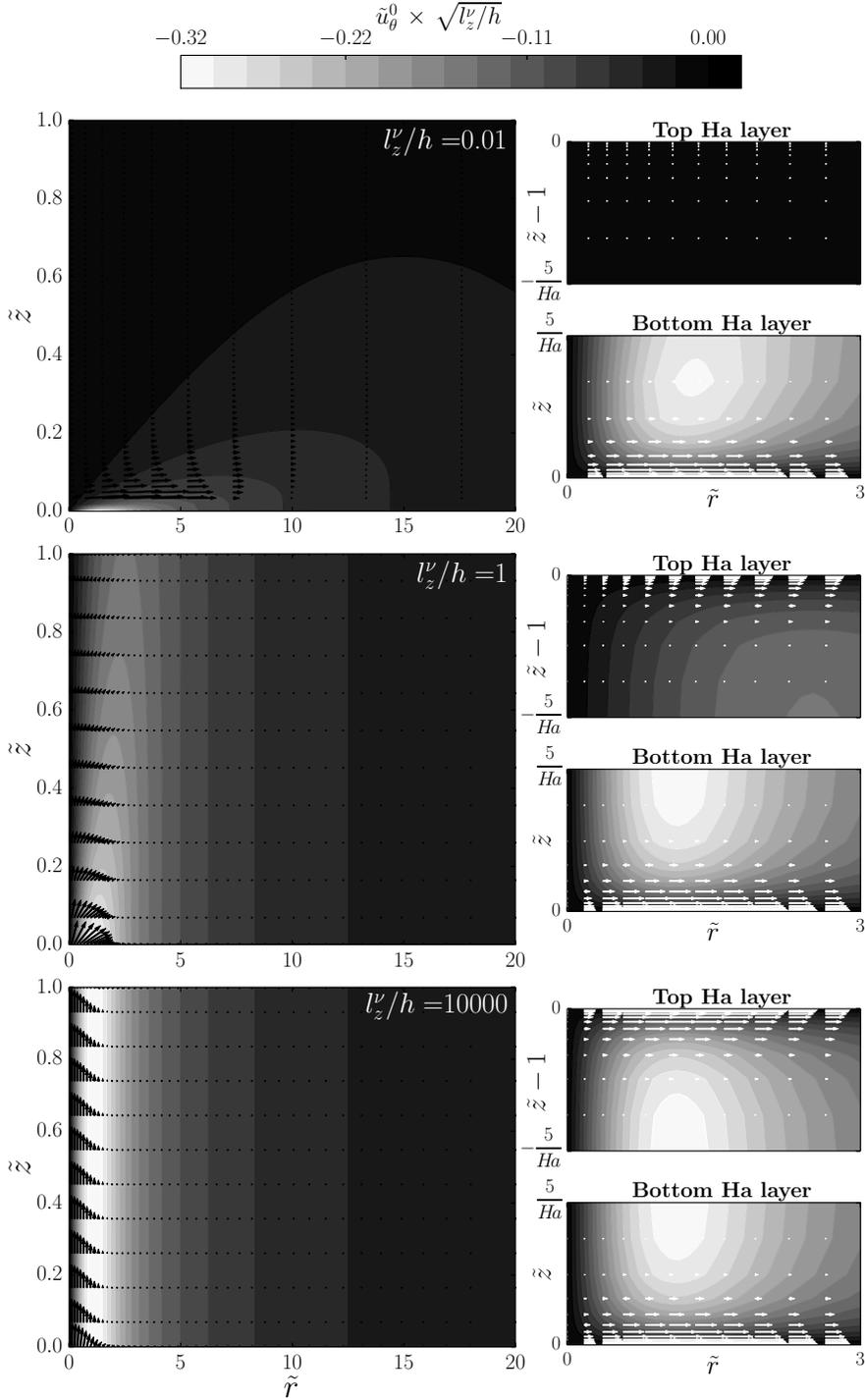

Figure 5: solution at leading order for $\underline{Ha} = 3644$, and for $l_z^\nu/h = 0.01$, 1 and 10000 respectively. The magnitude of $\tilde{u}_\theta^0 \times \sqrt{l_z^\nu/h}$ is indicated by filled contours. The electric current density is represented by black and white vectors (color choice is just a matter of contrast). Scaling of vectors has been adapted to compensate the much higher current densities in the boundary layers. Inserts: close up view of the top and bottom Hartmann layers featuring the usual exponential profile for the electric current.

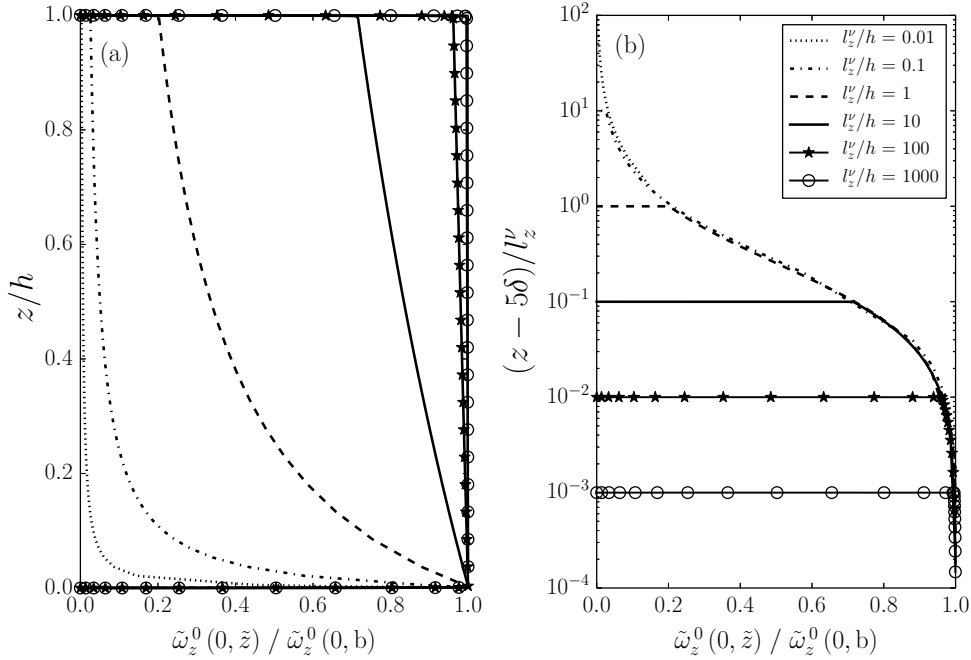

Figure 6: axial vorticity $\tilde{\omega}_z^0(0, \tilde{z})$ normalized by the vorticity right outside the bottom Hartmann layer $\tilde{\omega}_z^0(0, b)$ for $Ha = 3644$. (a): $z$ is normalized by the height of the channel $h$. (b): $z$ is normalized by the Lorentz force diffusion length $l_z^\nu$.

Close up views of the Hartmann layers are given in the inserts of figure 5. It is clear from these figures that the electric content of all bottom Hartmann layers is quite similar, but that the electric content of the top Hartmann layer depends on how far the Lorentz force is able to diffuse momentum along $\tilde{z}$. As expected, the electric current decreases away from the walls following an exponential profile in all cases.

The dimensionality of the base flow is better quantified with figure 6. From now on, $\tilde{\omega}_z^0(\tilde{r}, t)$ and $\tilde{\omega}_z^0(\tilde{r}, b)$ refer to the vorticity right outside the top and bottom Hartmann layers respectively (see Pothérat *et al.* (2002) for a mathematically rigorous definition of this concept). Figure 6 portrays the profile of vertical vorticity $\tilde{\omega}_z^0(0, \tilde{z})$ normalized by $\tilde{\omega}_z^0(0, b)$ along the axis of the channel. In figure 6.a, all structures evolve in a channel of fixed height ($\tilde{z}$ is normalized by $h$). This representation highlights the effect of the ratio $l_z^\nu/h$ on the dimensionality of the base flow: as $l_z^\nu/h$ increases, the momentum induced right above the injection electrode is diffused farther and farther by the Lorentz force, hence progressively smoothing out velocity gradients along $\tilde{z}$. In figure 6.b, all curves are shifted down by $5\delta$ to account for the varying thickness of the Hartmann layer, and then normalized by $l_z^\nu$. The collapse of all curves in these variables clearly indicates that all vortices follow a universal profile, which is solely defined by the competition between the Lorentz force and viscous dissipation. In other words, the effect of the vertical confinement is local, and only consists in ending the universal profile by introducing a no-slip boundary (the presence of the top wall is felt over a distance whose order of magnitude

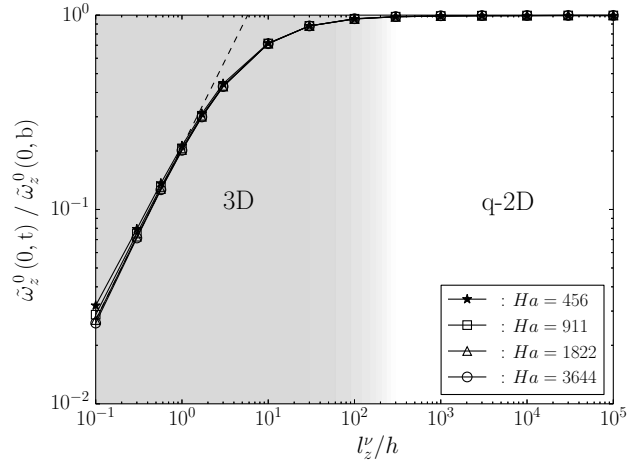

Figure 7: dimensionality of the base flow. 3D when $\bar{\omega}_z^0(0,t)/\bar{\omega}_z^0(0,b) < 1$ ; quasi-2D when $\bar{\omega}_z^0(0,t)/\bar{\omega}_z^0(0,b) \to 1$. A color gradient is used to highlight the smooth transition from a 3D to a quasi-2D base flow as the parameter $l_z^\nu/h$ increases.

is no larger than the thickness of the Hartmann layer).

Figure 7 compares the vorticity on the axis of the channel right below the top Hartmann layer to the vorticity right above the bottom one. This figure highlights how all the information about the dimensionality of the base flow is exclusively contained within the single parameter $l_z^\nu/h$, regardless of the value of the Hartmann number. Furthermore, the transition from 3D to quasi-2D base flows appears to be a gradual process that occurs around $l_z^\nu/h \sim 100$. This effect was also noted by Klein & Pothérat (2010) in turbulent flows.

### 6.2. The topology of meridional recirculations

Figures 8 and 9 give a complete view of the velocity field for the low- and high- $Ha$ cases. When the base flow is 3D ($l_z^\nu/h < 100$), a large counter-clockwise recirculation dominates the flow. This phenomenon has been observed by Akkermans et al. (2008) in electrolytes, and by Pothérat et al. (2013) in steady and turbulent liquid metal flows. It is driven by an axial pressure gradient that builds up along the axis of the vortex as a result of the negative gradient of azimuthal velocity along $\mathbf{e}_z$. Since the associated flow recirculates in opposite direction to Ekman pumping, and that both are driven by differential rotation, it was called inverse Ekman pumping by the latter authors. As the base flow becomes increasingly quasi-2D ($l_z^\nu/h > 100$), a clockwise recirculation becomes visible at the bottom of the domain, and grows steadily with $l_z^\nu/h$. The secondary flow is then composed of two counter-rotating structures, which correspond to direct Ekman pumping (Ekman 1905), or what is also called the "tea-cup effect". Unlike inverse pumping (which stems from a pressure gradient along the axis of the vortex), direct pumping is driven by a radial pressure gradient inside the boundary layers, which develops in the bulk to oppose centrifugal forces. To further elucidate the role of pressure gradients on the topology of secondary flows, let us reconstruct numerically from the Navier-Stokes equation projected along $\mathbf{e}_z$ the vertical pressure gradient along the axis of the vortex:

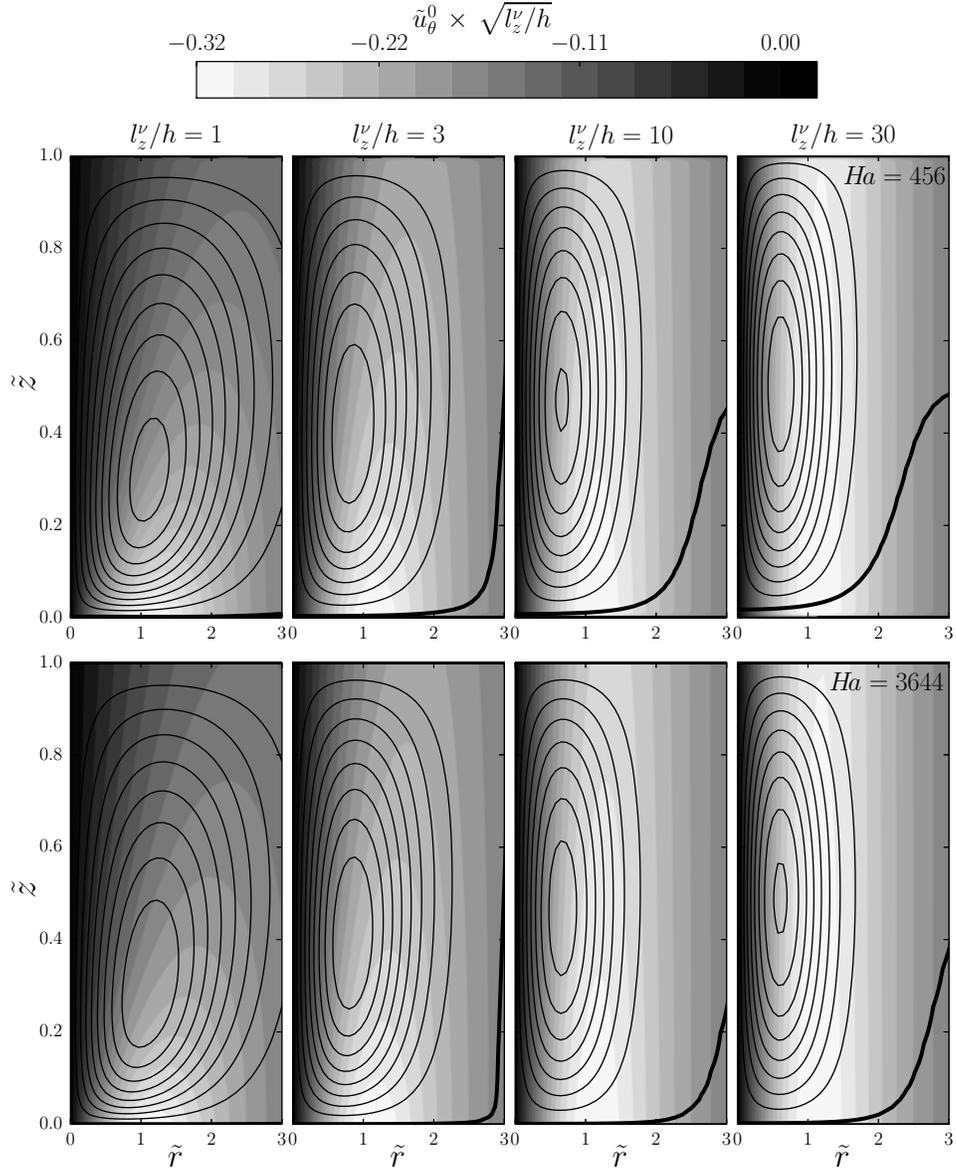

Figure 8: complete velocity field for 3D base flows. Top: $Ha = 456$; Bottom: $Ha = 3644$. The magnitude of $\tilde{u}_\theta^0$ is indicated by filled contours. Streamlines correspond to iso-values of $\psi^1$. ━━: counter-clockwise recirculation ($\psi^1 < 0$), ━━: $\psi^1 = 0$.

$$\frac{\partial \tilde{p}^0}{\partial \tilde{z}} = \frac{2}{Ha} \left(\frac{l_z^\nu}{h}\right)^{-1} \frac{\partial^2 \tilde{u}_{\tilde{z}}^1}{\partial \tilde{r}^2}\bigg|_{0,\tilde{z}} + \frac{1}{Ha^2} \frac{\partial^2 \tilde{u}_{\tilde{z}}^1}{\partial \tilde{z}^2}\bigg|_{0,\tilde{z}}. \qquad (6.1)$$

The profiles of pressure gradient along the axis of the vortex are represented in figure 10

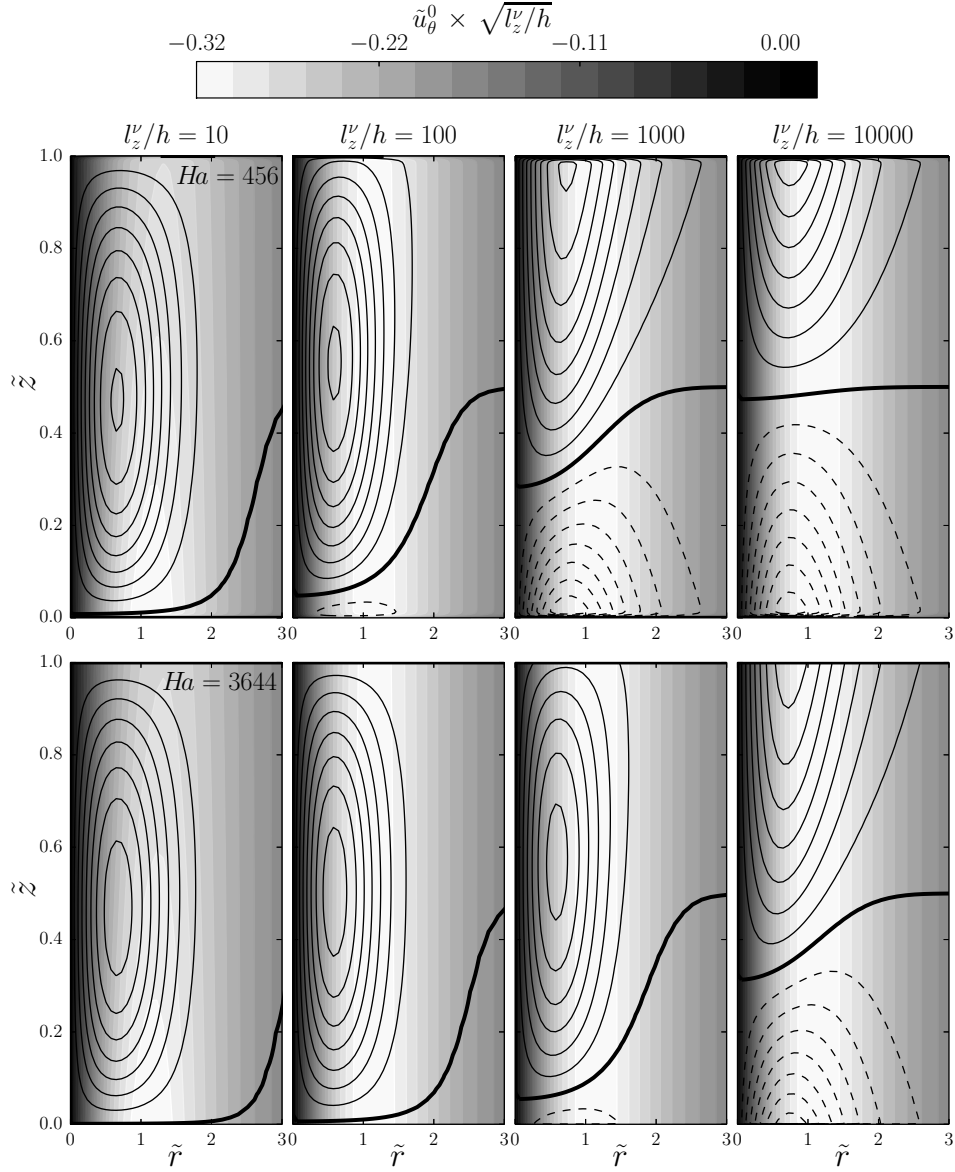

Figure 9: complete velocity field for increasingly quasi-2D base flows. Top: $Ha = 456$; Bottom: $Ha = 3644$. The magnitude of $\tilde{u}_\theta^0$ is indicated by filled contours. Streamlines correspond to iso-values of $\psi^1$. **- - -**: clock-wise recirculation ($\psi^1 > 0$), **——**: counter-clockwise recirculation ($\psi^1 < 0$), **——**: $\psi^1 = 0$.

in order to illustrate the previous argument. When the flow is 3D, a positive pressure gradient exists in the bulk, whose effect is to drive a jet down along the axis of the vortex. Because this phenomenon is entirely governed by velocity gradients in the core, it is no surprise that the intensity of the inverse pumping is driven by $l_z^\nu/h$. As a result of quasi-two dimensionality, the dependence of the pressure (or any other quantity for that matter) on $\tilde{z}$ in the bulk disappears. However, a very strong vertical pressure gradient exists at both ends of the axis as a result of a converging radial flow within the boundary layers.

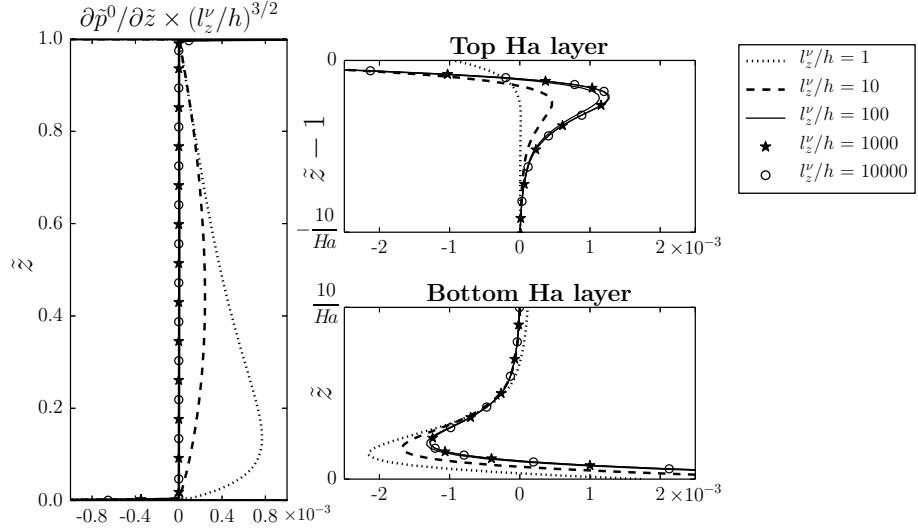

Figure 10: pressure gradient along the axis of the vortex. 3D base flows introduce a positive pressure gradient along $\mathbf{e}_z$ in the bulk, which drives a jet to flow down the axis of the channel. In the Hartmann layers, the negative pressure gradient at the bottom pushes the fluid up along the axis, while the positive pressure gradient at the top pulls it down.

Interestingly, a negative pressure gradient always exists in the bottom Hartmann layer regardless of whether the base flow is 3D or quasi-2D. This means that a recirculation always exists at the bottom of the channel (though it is not always visible), which results from direct pumping. By contrast, a positive pressure gradient does not exist in the top Hartmann layer for $l_z^\nu/h = 1$, meaning that in this particular case, the top recirculation is exclusively driven by inverse pumping due to the vertical pressure gradient in the lower half of the channel. As a matter of fact, figure 10 showcases the progressive shift in the mechanism driving the top recirculation, which is not obvious a priori, as it does not transpire in its topology. To summarize, direct and inverse pumping co-exist in all cases investigated.

Figure 11 gives a close up view of the bottom Hartmann layers for the smaller values of $l_z^\nu/h$. This figure confirms the existence of a weak clockwise recirculation in the Hartmann layers, although this direct recirculation is too weak to balance the downwards axial jet. The confinement of this weak direct pumping to the thin boundary layers makes it very difficult to fully capture whether experimentally or numerically. Yet, it is a clear feature of the analytical solution. In addition, figure 11 demonstrates that the Hartmann number does not actually impact the mechanisms driving the meridional flow, but modifies its topology instead by squeezing the streamlines closer to the walls. In order to further

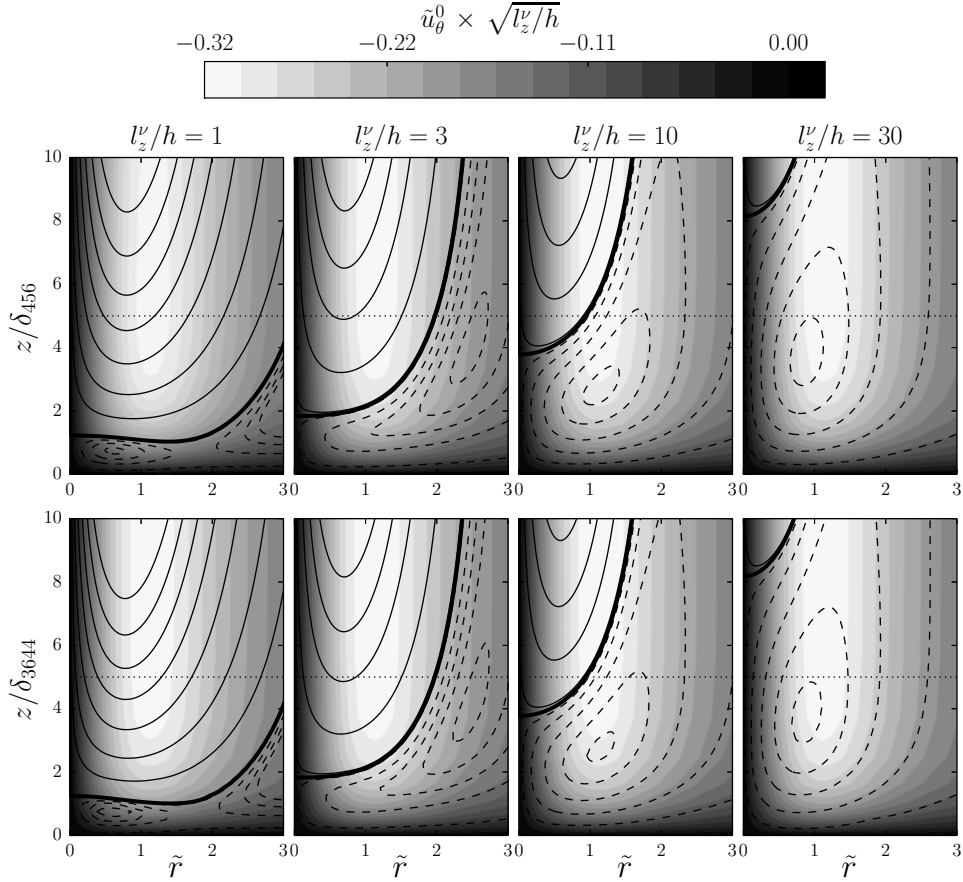

Figure 11: close up view of the velocity field in the bottom Hartmann layer for 3D base flows. Top: $Ha = 456$; Bottom: $Ha = 3644$. The magnitude of $\tilde{u}_\theta^0$ is indicated by filled contours. Streamlines correspond to iso-values of $\psi^1$. - - -: clock-wise recirculation ($\psi^1 > 0$), —: counter-clockwise recirculation ($\psi^1 < 0$), —: $\psi^1 = 0$. ·····: plausible positioning of the outer edge of the Hartmann layer located at $z/\delta_{Ha} = 5$, where $\delta_{Ha} = h/Ha$ for $Ha = 456$ and $Ha = 3644$ respectively.

quantify the secondary flows, we introduce the poloidal flowrate $q_{\tilde{z}}^1(\tilde{z})$ defined by:

$$q_{\tilde{z}}^1(\tilde{z}) = \frac{1}{2} \int\limits_0^{2\pi} \int\limits_0^{\tilde{R}} |\tilde{u}_{\tilde{z}}^1(\tilde{r}, \tilde{z})| \, \tilde{r} \, d\tilde{r} \, d\theta. \qquad (6.2)$$

We also introduce $\tilde{h}_c = h_c/h$, where $h_c$ represents the height of the bottom recirculation. $\tilde{h}_c$ is found as the first local minimum of $q_{\tilde{z}}^1(\tilde{z})$. Figure 12 represents the height of the bottom recirculation against the variable $l_z^\nu/h \times Ha^{-1}$, which can also be interpreted as the ratio $(\eta/h)^2$ by virtue of (2.2). When the base flow is quasi two-dimensional ($l_z^\nu/h > 100$), the quantity $\eta/h$ may naturally be confused with the aspect ratio of the vortex. It appears from figure 12 that when the leading order is close to being quasi-

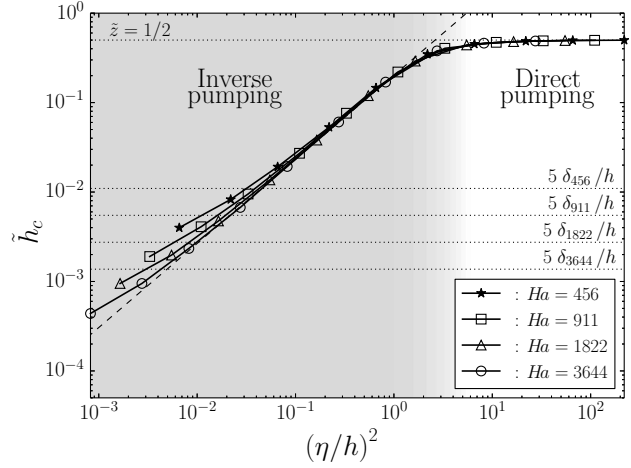

Figure 12: height of the bottom recirculation against $(\eta/h)^2$. The middle of the channel is located by $\tilde{z} = 1/2$. An estimation of the thickness of the bottom Hartmann layer is given by $\tilde{z} = 5\,\delta_{Ha}/h$, where $\delta_{Ha} = h/Ha$ for $Ha = 456,\ 911,\ 1822$ and $3644$ respectively.

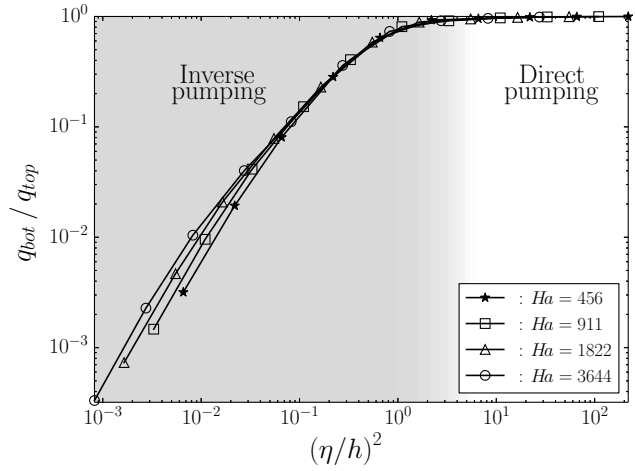

Figure 13: ratio between $q_{bot}$ and $q_{top}$ as a function of $(\eta/h)^2$ for $Ha = 456,\ 911,\ 1822$ and $3644$.

2D, the topology of the meridional flow is fully determined by the parameter $\eta/h$. More specifically, narrow aspect ratios correspond to inverse Ekman pumping, while wide aspect ratios lead to an asymptotic state where two counter rotating structures of equal size split the channel in half. The shift from inverse to direct pumping concurs with a ratio $(\eta/h)^2$ that is of order unity. Note that for a perfectly quasi-2D flow ($l_z^{\nu}/h \to \infty$) only *direct* Ekman pumping subsists for any finite value of $\eta/h$.

In order to quantify the relative intensity of secondary flows, we introduce $q_{bot}$ and $q_{top}$ the mean flowrates occurring inside the bottom and top recirculation, respectively

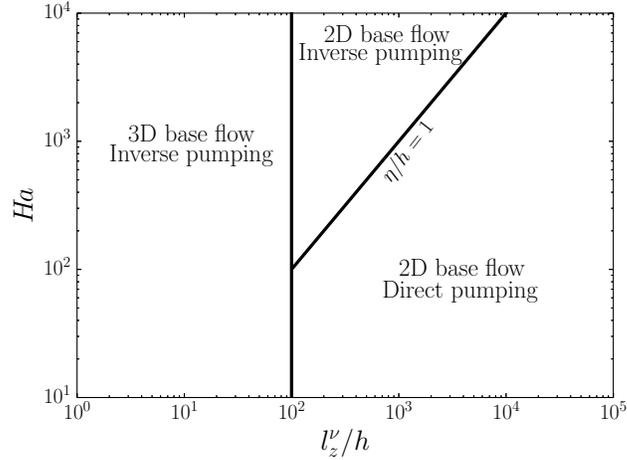

Figure 14: phase diagram summarizing all features of wall bounded low-Rm MHD vortices.

defined as:

$$q_{bot} = \frac{1}{\tilde{h}_c} \int\limits_0^{\tilde{h}_c} q_{\tilde{z}}^1(\tilde{z}) \; d\tilde{z} \qquad \text{and} \qquad q_{top} = \frac{1}{1 - \tilde{h}_c} \int\limits_{\tilde{h}_c}^1 q_{\tilde{z}}^1(\tilde{z}) \; d\tilde{z}. \qquad (6.3)$$

According to figure 13, the ratio $q_{bot}/q_{top}$ can also be fully described by the unique variable $(\eta/h)^2$ when the flow is close to being quasi-2D. More specifically, Ekman pumping starts dominating the flow when $(\eta/h)^2$ increases beyond unity, at which point the intensity of both recirculations converges towards the same value.

The behavior of $\tilde{h}_c$ and $q_{bot}/q_{top}$ are quite similar, which leads to a robust criterion expressing whether the secondary flow is driven by inverse or direct pumping, namely: $(\eta/h)^2 > 1$. A phase diagram summarizing all different configurations is reported in figure 14. More specifically, it underlines the fact that inverse pumping can still exist when the base flow is close to being 2D, if the vortex is of sufficiently small aspect ratio $\eta/h$. This comes from the very nature of direct pumping, which originates within the boundary layers and is therefore strongest there. In thin vortices, its influence on the bulk is limited, whereas a small pressure gradient in the bulk suffices to drive inverse pumping.

### 6.3. Is two-dimensionality a good source of helicity?

Having now characterized both the azimuthal and meridional flows, we are in a position to determine their potential to generate helicity. Figure 15 displays the helicity $\tilde{H}(\tilde{z})$ integrated over a cross-section of height $\tilde{z}$:

$$\tilde{H}(\tilde{z}) = 2\pi \int\limits_0^{\tilde{R}} \tilde{\mathbf{u}} \cdot \tilde{\boldsymbol{\omega}} \; \tilde{r} d\tilde{r}, \qquad (6.4)$$

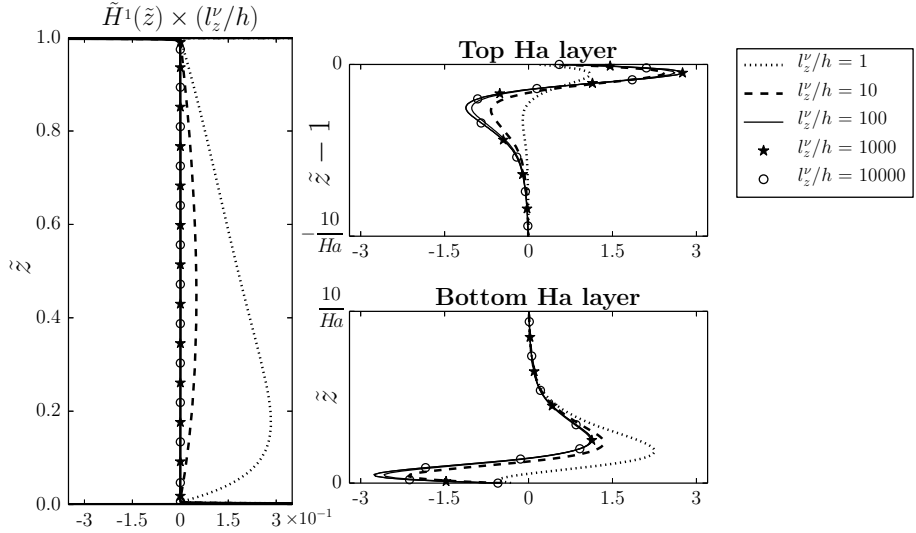

Figure 15: local helicity integrated over horizontal cross sections $H^1(\tilde{z})$.

which happens to be significant only at first order i.e. $\tilde{H}(\tilde{z}) = N^{-1}\tilde{H}^1 + O(N^{-3})$, with

$$\tilde{H}^1(\tilde{z}) = 2\pi \int\limits_0^{\tilde{R}} \left( \tilde{u}_\theta^0 \tilde{\omega}_\theta^1 + \tilde{u}_{\tilde{r}}^1 \tilde{\omega}_{\tilde{r}}^0 + \tilde{u}_{\tilde{z}}^1 \tilde{\omega}_{\tilde{z}}^0 \right) \tilde{r}d\tilde{r}. \qquad (6.5)$$

Figure 15 suggests that helicity exists in the bulk when the base flow is 3D, while quasi two-dimensionality confines helicity to the boundary layers. Furthermore, $\tilde{H}^1(\tilde{z})$ is non symmetrical only for 3D base flows, meaning that the global production of helicity will be non zero only in this case. Consequently, Ekman pumping does not appear to be the most favorable source of global helicity in MHD vortices, which may seem surprising at first. It can however be attributed to several factors. First, secondary flows are a lot stronger when inverse pumping dominates. Second, Ekman pumping introduces symmetrical structures of opposite sign, which compensate each other globally. Helicity generated by inverse pumping on the other hand conserves its sign across the entire layer.

Finally, direct and inverse pumping appear to produce helicity in different ways. While inverse Ekman pumping collocates the axial velocity and vorticity in the bulk, direct pumping produces helicity mainly by combining centripetal jets and shear within the boundary layers.

## 7. Conclusion

We showed in this paper that the topology of an electrically driven vortex confined between two no-slip walls in the low-Rm approximation and weakly inertial limit can be fully described with two parameters. On the one hand, the dimensionality of the leading order is uniquely characterized by the ratio $l_z^\nu/h$, which compares the range of action of the Lorentz force to the height of the channel. On the other hand, the topology of the secondary recirculations is fully described by the parameter $\eta/h$, which compares the width of the injection electrode to the height of the channel. In the quasi two-dimensional limit, $\eta/h$ may be interpreted as the aspect ratio of the vortex.

Thanks to the analytical approach that was undertaken, we were able to completely resolve the finest properties of the flow and in particular the Hartmann boundary layers, which are an inherent source of three-dimensionality. The Hartmann numbers used in this study were comparable to those found experimentally. Using this approach, we were able to distinguish two different inertial mechanism able to drive the first order recirculations: inverse and direct Ekman pumping. We found out that both co-existed in all cases investigated (although direct pumping is confined to the bottom boundary layer when the base flow is three-dimensional), and that the shift from one mechanism to the other occurred smoothly. This result could not have been obtained either numerically or experimentally due to a lack of resolution of either approaches at high $Ha$.

Finally, it was found that global helicity is expected to be prominent only in three-dimensional configurations. In a geophysical context, this result might help clarify the question of whether Ekman pumping is a relevant source of helicity to sustain planetary dynamos. As noted by Davidson (2014), Eckman pumping may not be a very efficient source of helicity in planetary cores because quasi-2D vortices extending across the liquid core of the Earth for example are unlikely to exist. They can therefore be expected to be three-dimensional, hence favoring alternative mechanisms such as inverse pumping or the propagation of inertial waves.

The authors are grateful to the CNRS laboratories LNCMI and CRETA, and in particular to François Debray and André Sulpice for hosting and supporting their work. The laboratory SIMAP is part of the LabEx Tec 21 (Investissements d'Avenir - Grant Agreement No. ANR-11-LABX-0030).